\documentclass[9pt,twocolumn,twoside]{pnas-new}

\usepackage{xcolor}

\newcommand{\MDGrevise}[1]{{{\color{black}#1}}}
\newcommand{\DFrevise}[1]{{{\color{black}#1}}}

\templatetype{pnasresearcharticle} 

\usepackage{natbib}
\title{Discovering multiscale and self-similar structure with data-driven wavelets}

\author[a]{Daniel Floryan}
\author[a,1]{Michael D. Graham} 

\affil[a]{Department of Chemical and Biological Engineering, University of Wisconsin--Madison, Madison, WI 53703}

\leadauthor{Floryan} 

\significancestatement{Multiscale structure is all around us: in biological tissues, active matter, oceans, networks, and images. Identifying the multiscale features of these systems is crucial to our understanding and control of them. We introduce a method that rationally extracts localized multiscale features from data, which may be thought of as the building blocks of the underlying phenomena. }

\authorcontributions{D.F. and M.D.G. conceptualized the work; D.F. developed the DDWD formulation and algorithm, and generated the computational results; D.F. and M.D.G. designed the computational study, analyzed the results, and wrote the paper. }
\authordeclaration{The authors declare no conflicts of interest.}
\correspondingauthor{\textsuperscript{1}To whom correspondence should be addressed. E-mail: mdgraham@wisc.edu}

\keywords{wavelet $|$ multiscale $|$ data-driven decomposition $|$ machine learning $|$ turbulence} 

\begin{abstract}

Many materials, processes, and structures in science and engineering have important features at multiple scales of time and/or space; examples include biological tissues, active matter, oceans, networks, and images. 
Explicitly extracting, describing, and defining such features are difficult tasks, at least in part because each system has a unique set of features.
Here, we introduce an analysis method that, given a set of observations, discovers an energetic hierarchy of structures localized in scale and space. We call the resulting basis vectors a ``data-driven wavelet decomposition''. We show that this decomposition reflects the inherent structure of the dataset it acts on, whether it has no structure, structure dominated by a single scale, or structure on a hierarchy of scales. In particular, when applied to turbulence---a high-dimensional, nonlinear, multiscale process---the method reveals self-similar structure over a wide range of  spatial scales, providing direct, model-free evidence for a century-old phenomenological picture of turbulence. This approach is a starting point for the characterization of localized hierarchical structures in multiscale systems, which we may think of as the building blocks of these systems. 
\end{abstract}

\dates{This manuscript was compiled on \today}
\doi{\url{www.pnas.org/cgi/doi/10.1073/pnas.XXXXXXXXXX}}

\begin{document}

\maketitle
\thispagestyle{firststyle}
\ifthenelse{\boolean{shortarticle}}{\ifthenelse{\boolean{singlecolumn}}{\abscontentformatted}{\abscontent}}{}

\dropcap{M}any important
processes are multiscale in nature, meaning that they exhibit structure at multiple scales of time and/or space. In nature, a prominent example is the dynamics of oceans and associated interactions with the atmosphere, which govern the planet's weather and climate systems \cite{lau2011intraseasonal}; much effort is expended in capturing and understanding effects at multiple scales of time and space \cite{lermusiaux2013multiscale}. In engineering, a prominent example is networks, specifically social media networks. Networks have multiscale structure by virtue of hierarchies of communities of nodes in the networks \cite{ahn2010link}. Understanding the structure of hierarchical communities in social media networks is crucial to understanding the spread of disinformation (and censorship of information) in these networks \cite{bradshaw2018challenging}. Broadly speaking, identifying and understanding the features present in multiscale processes is crucial to understanding and controlling these processes. 
Although the application we focus on here will be turbulent fluid flows, the ensuing discussion applies to any multiscale process for which the notions of energy (variance in the statistical context) and localization (a form of sparsity) are relevant. 

\begin{figure}
  \begin{center}
  \includegraphics[width=\linewidth]{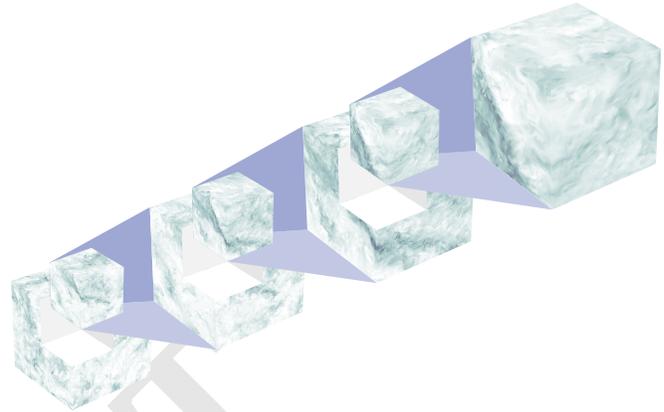}
  \end{center}
  \caption{Snapshot of homogeneous isotropic turbulence from the Johns Hopkins Turbulence Database \citep{perlman2007data,li2008public,yeung2012dissipation}, showing the kinetic energy per unit mass, with darker colour corresponding to greater energy. }
  \label{fig:multiscale}
\end{figure}

 Turbulence is a canonical multiscale process consisting of localized concentrations of vortex motion that are coherent in space and time and coexist at a wide range of scales. Theoretical arguments indicate that at intermediate scales and far from walls, the structure of a turbulent flow should be self-similar \cite{Marusic:2019fy,Sagaut:2018jh}. This notion is qualitatively illustrated in Figure~\ref{fig:multiscale}, which illustrates a snapshot from a simulation of homogeneous isotropic turbulence (HIT) at several scales \footnote{{http://turbulence.pha.jhu.edu}}\citep{perlman2007data,li2008public,yeung2012dissipation}. As with other multiscale processes, a great challenge in fluid dynamics is to rationally identify and analyze coherent structures from a complex turbulent flow field.  While it is often mathematically convenient to analyze signals in the Fourier domain, trigonometric functions are not localized in space, and what one observes at an instant in time in a turbulent flow rarely if ever looks sinusoidal. Alternately, conventional wavelet bases, which are localized and self-similar, can be used for analysis \cite{farge1992wavelet}. In both the Fourier and wavelet approaches, the bases for representing the flow  are imposed a priori rather than emerging from data.
 

 One of the primary methods of extracting structure from data is principal components analysis (PCA), which in fluid dynamics is typically denoted Proper Orthogonal Decomposition (POD) \cite{Holmes:2012gw}.
 (See \cite{taira2017modal} for other popular modal decomposition methods.) Given an ensemble (often a time series) of data, PCA yields a data-driven orthogonal basis whose elements are optimally ordered by energy content. When applied to velocity field data for a fluid flow, the resulting basis elements may be thought of as the building blocks of that flow, and its application has yielded many structural and dynamical insights \cite{Holmes:2012gw,Hellstrom:2016dw}. 
 One limitation of PCA is that the  basis elements tend not to be localized in space; indeed,  for directions in which a field is statistically homogeneous, the PCA basis elements are Fourier modes \cite{Holmes:2012gw}. In this case, not only do the PCA modes have no localization in space, they also reveal no information about the flow beyond what Fourier decomposition would provide. 

A well known formalism that produces bases with spatially localized elements is that of wavelets. The name is quite descriptive: wavelets are localized waves. In particular, wavelet decompositions  provide an orthogonal basis whose elements are localized in both space and scale. Traditionally, the basis elements are translations and dilations of a single vector called the mother wavelet \cite{STRANG:1989vn,daubechies1992ten, meyer1992wavelets, mallat1999wavelet, frazier2006introduction}. The Supplementary Information provides a concise summary of results relevant to the present work. Traditional wavelet methods (where the mother wavelet is prescribed a priori) have already found use in turbulence precisely because of the space-scale unfolding they produce \cite{argoul1989wavelet, everson1990wavelet, farge1992wavelet, farge2001coherent, farge2003coherent, okamoto2007coherent, ruppert2009wavelet, yamada1991identification, meneveau1991analysis}, giving hope that data-driven methods based on wavelets may lead to new insights into turbulence.

A myriad of data-driven methods of structure identification and extraction based on wavelets have been developed (e.g., \cite{gilles2013empirical, gopinath1994optimal, grasemann2004evolving, mendez2019multi, ophir2011multi, recoskie2018alearning, recoskie2018blearning, recoskie2018gradient, sogaard2017learning, tewfik1992optimal, zhuang1994optimal}). Although these methods may yield localized structures, they are limited in that the construction of the resulting basis elements is prescribed in either scale or frequency, and many impose self-similarity on the basis, as is done with traditional wavelets. 
(The ``empirical wavelet transform''  of \cite{gilles2013empirical} does not have this feature, but relies on the existence of local maxima in the power spectrum of a signal, making it ill-suited to phenomena like turbulence without such local maxima.) 

In the present work we develop a method that integrates the data- and energy-driven nature of PCA with the space and scale localization properties of wavelets.
As our derivation and illustrative examples will reveal, we impose very little structure in our method, so any structure in the basis may be attributed to the underlying structure of the dataset under consideration. We call the resulting basis a ``data-driven wavelet decomposition'' (DDWD), and use it to gain insights into the structure of turbulence, though we emphasize that the method is general in its application.

\section*{Formulation}

Before presenting the DDWD, it will be useful to introduce key features of PCA and conventional wavelet decompositions.
Suppose we have a dataset $\{z_i\}_{i=1}^M \in \mathbb{R}^N$,  each $z_i$ being a sample data vector (e.g., one component of a velocity field uniformly sampled along a line through the flow). We can arrange the dataset into a matrix $Z \in \mathbb{R}^{N \times M}$ whose columns are the data vectors $z_i$, normalized so that $\mathrm{tr}\:ZZ^T=1$ (the normalization does not change the results of PCA, but is done here because it parallels our formulation of DDWD later). PCA seeks an ordered orthonormal basis $\{ \phi_i \}_{i = 1}^N$ such that the energy of the dataset projected onto the first $K\leq N$ basis elements is maximized. One way to state this problem, which parallels our later description of data-driven wavelets, is as follows. We determine the first basis element $\phi_1$ so that the projection of the data onto this element is maximized. This problem can be written
\begin{eqnarray}
  \label{eq:1}
  \max_{\phi} \quad & \phi^T ZZ^T\phi \\
  \textrm{s.t.} \quad & \phi^T \phi = 1.
\end{eqnarray}
The solution to this problem is the eigenvector of $ZZ^T$ with the largest eigenvalue. The second basis element $\phi_2$ is found by projecting out the component of the data in the $\phi_1$ direction and repeating, yielding that $\phi_2$ is the eigenvector of $ZZ^T$ with the second largest eigenvalue. Basis elements $\phi_i$ solve
\begin{eqnarray}
  \label{eq:2}
  \max_{\phi} \quad & \left \| \phi^T \left(Z - \sum_{j = 1}^{i - 1} \phi_j \phi_j^T Z \right) \right \|_2^2 \\
  \textrm{s.t.} \quad & \phi^T \phi = 1, \quad \phi^T \phi_j = 0,\, j = 1, \ldots, i - 1.
\end{eqnarray}
This formulation is recursive, producing a hierarchy of subspaces ordered by how much of the dataset's energy (Frobenius norm) they contain: $\mathbb{R}^N = \text{span}\{\phi_1\} \oplus \ldots \oplus \text{span}\{\phi_N\}$. The basis elements $\phi_i$ are the eigenvectors of $ZZ^T$. For statistically homogeneous data in a periodic domain, $ZZ^T$ (more precisely, its expected value) is circulant, in which case the $\phi_i$ are simply discrete Fourier modes. 

\begin{figure}[t]
  \begin{center}
  \includegraphics[width=\linewidth]{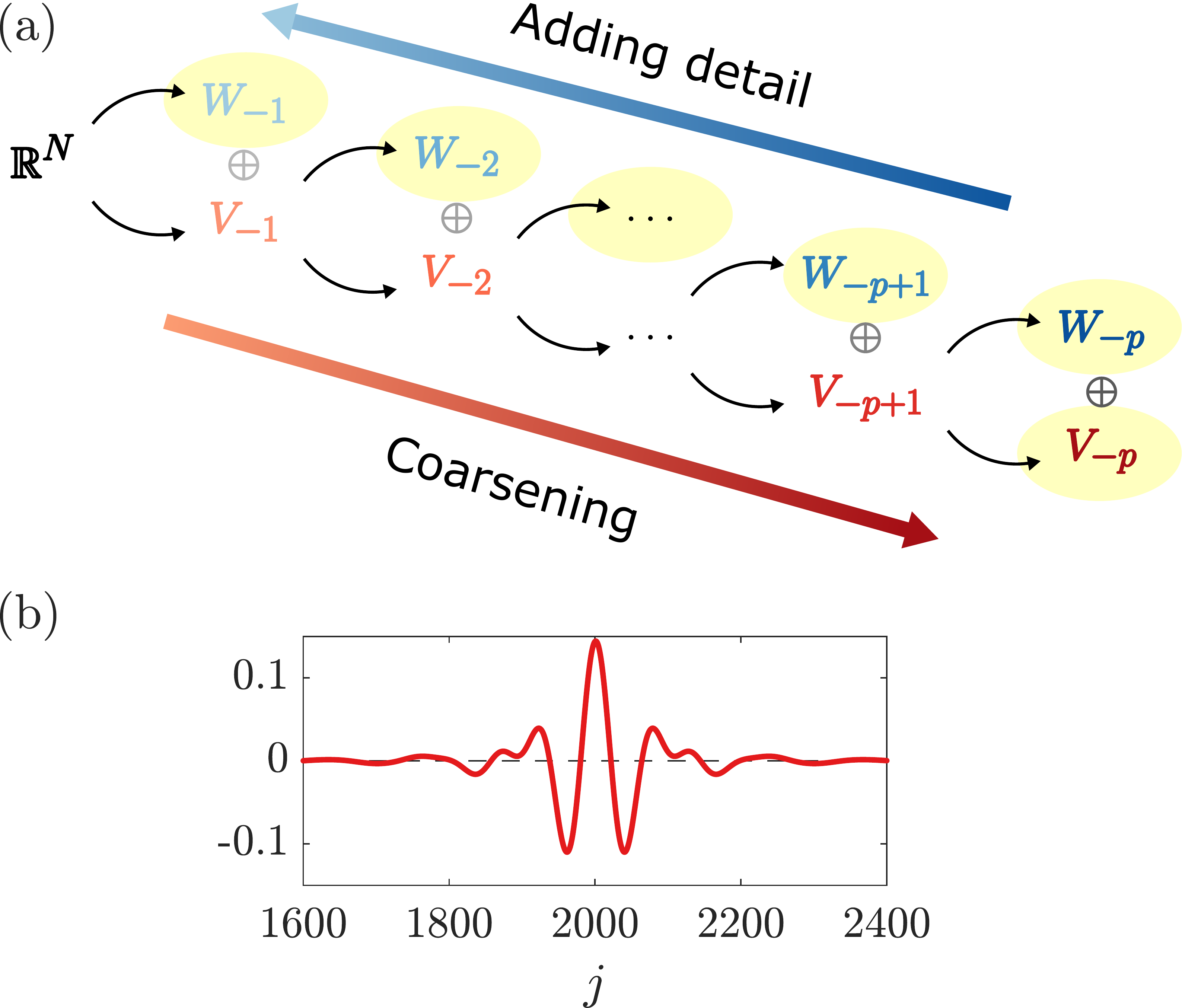}
  \end{center}
  \caption{(a) Subspaces from wavelets on $\mathbb{R}^N$. At stage $l$, approximation subspace $V_{-l}$ is split into detail subspace $W_{-l-1}$ and approximation subspace $V_{-l-1}$, each half the dimension of $V_{-l}$. Subspace $V_{-l}$ is spanned by the $N/2^l$ translates by $2^l$ of $\phi_{-l}$, and $W_{-l}$ is spanned by the $N/2^l$ translates by $2^l$ of $\psi_{-l}$. The full space is decomposed into progressively coarser subspaces, $\mathbb{R}^N = W_{-1} \oplus \ldots \oplus W_{-p} \oplus V_{-p}$, or, going the other way, into the addition of progressively finer details. These subspaces are highlighted. In the present work, an ensemble of data is used to define a specific decomposition of this form. (b) Discrete Meyer wavelet for $N = 4096$ and $l = 6$. }
  \label{fig:waveSubspaces}
\end{figure}

Traditional wavelet decompositions also produce a hierarchy of orthogonal subspaces, but there are important differences from PCA. First, the basis elements are not determined from data, but are selected a priori; there are many standard options \citep{mallat1999wavelet}. Second, by construction, the decomposition produces  
a hierarchy of orthogonal subspaces ordered by scale, as shown in Figure~\ref{fig:waveSubspaces}(a). We consider periodic vectors on $\mathbb{R}^N$, with $N$ even \cite{frazier2006introduction}. This space is split into subspaces $V_{-1}$ and $W_{-1}$, each of dimension $N/2$, and each spanned by the even translates of vectors $\phi_{-1}$ (the father wavelet) and $\psi_{-1}$ (the mother wavelet), respectively. Once $\phi_{-1}$ is known, $\psi_{-1}$ can be found, and vice versa. 
The father and mother wavelets, and their even translates, are mutually orthonormal by construction. Subspace $V_{-1}$ is called an approximation subspace because it contains all the low frequencies, and $W_{-1}$ is called a detail subspace because it contains all the high frequencies. Given a signal, its projection onto $V_{-1}$ produces a low-pass filtered version of the signal, and its projection onto $W_{-1}$ produces the detail needed to reconstruct the full signal. We then recursively split the approximation subspaces. For $N = 2^p$ (which we assume throughout), we get a hierarchy of subspaces of progressively coarser scales: $\mathbb{R}^N = W_{-1} \oplus \ldots \oplus W_{-p} \oplus V_{-p}$. For traditional wavelets, the sets of wavelets $\{\phi_i\}$ and $\{ \psi_i \}$ are determined from the father and mother wavelets, respectively, by a rescaling operation that is essentially a simple dilation \MDGrevise{by a factor of two} (see the Supplementary Information for more details). 
\MDGrevise{This process leads to a hierarchical basis structure of the form shown in Figure \ref{fig:waveSubspaces}(a).}

The DDWD combines the hierarchical structure of wavelets that is shown in Figure~\ref{fig:waveSubspaces}(a) with the energetic optimization of PCA. Namely, each time we split a subspace, we design the subsequent subspaces so that the approximation subspace contains as much of the dataset's energy as possible. 

The first step of the process is to find the \emph{wavelet generator} $u$, for which the projection of the data onto this vector and its even translates is maximized. We define $V_{-1}$ as the subspace spanned by these vectors, thus beginning the data-driven construction of a hierarchy with the structure of Figure~\ref{fig:waveSubspaces}(a).  This maximization is subject to (1) the constraint that $u$ and its even translates are mutually orthonormal, and (2) a penalty on the width of $u$, as measured by its circular variance $\mathrm{Var}(u)$; see the Supplementary Information for more details. This problem is stated as
\begin{eqnarray}
  \label{eq:3}
  \max_{u} & u^T A u - \lambda^2 \mathrm{Var}(u), \quad A = \frac{1}{\Vert Z \Vert_F^2} \sum_{k = 0}^{N/2 - 1} R^{-2k}ZZ^T R^{2k}\\
  \textrm{s.t.} & u^T R^{2k} u = \delta_{k0},\; k = 0, \ldots, N/2 - 1.
\end{eqnarray}
Here $\lambda$ measures the penalty on the variance, whose effect on the results we illustrate below, and $R$ is the circular shift operator: e.g.,~if $u=[a,b,c,d]^T$, then $Ru=[d,a,b,c]^T$. The solution $u$ and its even translates  generate the vectors $\phi_{-1}$ and $\psi_{-1}$; the former span $V_{-1}$ and the latter  $W_{-1}$. 
We then project the data onto $V_{-1}$,  replace $N$ by $N/2$ in the definition of $A$ and the orthonormality constraints, decrease $\lambda$ by a factor of $2$, and repeat, yielding $\phi_{-2}$ and $\psi_{-2}$, and thus the subspaces $V_{-2}$ and $W_{-2}$. We proceed recursively, finding the subspaces $V_{-l}$ and $W_{-l}$ such that $V_{-l}$ contains the maximal amount of energy of the dataset. 
Extensive details are found in the Supplementary Information. In the end, we find an energetic hierarchy of subspaces, optimized stage by stage, whose elements are orthogonal and localized. In contrast to previous data-driven methods incorporating wavelets, which impose restrictive structure, the only structure we impose is orthogonality, localization, and the hierarchy of Figure~\ref{fig:waveSubspaces}(a). 
  In the Supplementary Information, we also draw parallels between the DDWD and convolutional neural networks, and show how the DDWD naturally incorporates pooling and skip connections, two tricks that improve the performance of neural network architectures \cite{goodfellow2016deep}. Together with its inverse transform, the DDWD is akin to a convolutional autoencoder, but with the additional features of orthogonality of all elements, stage-wise energetic optimality, and the ability to unambiguously extract structure, which make the results interpretable. 

We make a point to note that for the DDWD, the stage $l$ of the hierarchy should not be conflated with the concept of scale. For traditional wavelets, stage and scale are interchangeable since whenever a subspace is split, the lower half of frequencies is always pushed to the approximation subspace and the upper half of frequencies is always pushed to the detail subspace. For the DDWD, however, the distribution of frequencies amongst the subspaces is dictated by energetic considerations, which depends on the dataset under consideration. An example below will elucidate this point. 

\section*{Results}

We will demonstrate the DDWD on three datasets with increasingly complex structure to show that the method extracts structure inherent to the data. 

\subsection*{Gaussian random data}

The first dataset we consider consists of Gaussian white noise, which has no structure. By construction, the basis produced by the DDWD is orthonormal, so the change-of-basis transformation is orthogonal. Any orthogonal transformation of Gaussian white noise produces Gaussian white noise. Therefore, in the absence of a variance penalty, applied to Gaussian white noise, the coordinates of the data in the DDWD basis (the wavelet coefficients) will be Gaussian white noise, so all wavelet coefficients will be uncorrelated and have variance equal to that of the input Gaussian white noise. That is, as long as we do not impose a variance penalty, this result implies that for Gaussian white noise there is no optimal set of wavelets, in the sense we have defined. In other words, the DDWD reflects that the dataset has no structure. If we do impose a variance penalty, then the optimal wavelets become discrete delta functions (i.e.,~the Euclidean basis vectors). The reason for this is simple: all wavelets capture the energy of white noise equally well, but the delta function will be the most localized among them.

\begin{figure}
  \begin{center}
  \includegraphics[width=\linewidth]{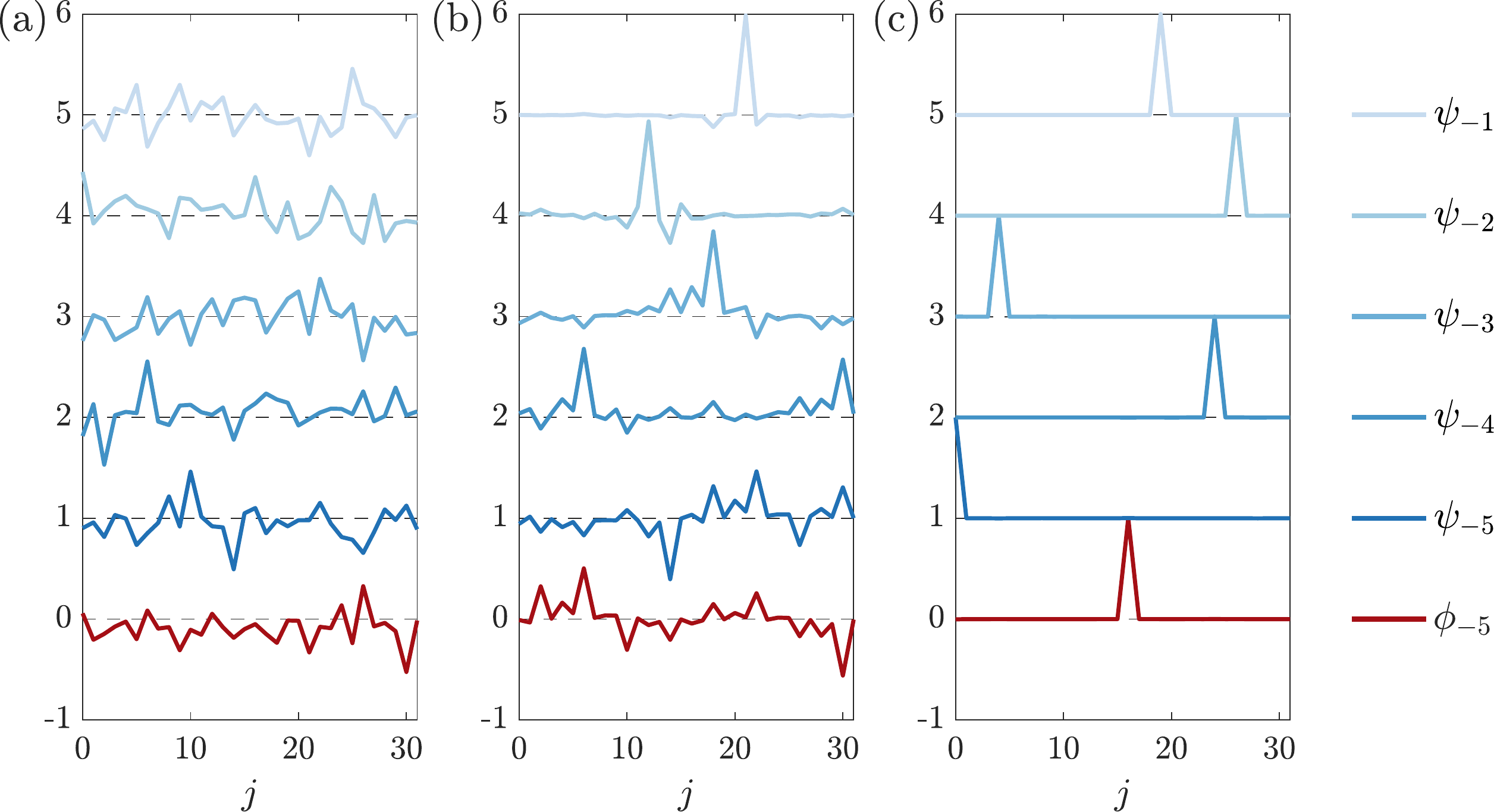}
  \end{center}
  \caption{White noise wavelets on $\mathbb{R}^{2^5}$. Colouring as in Figure~\ref{fig:waveSubspaces}(a). No variance penalty (a), small variance penalty (b), and large variance penalty (c). }
  \label{fig:randomWavelets}
\end{figure}

The result that all wavelets capture the energy of Gaussian white noise equally well highlights an interesting fact about the DDWD. In Figure~\ref{fig:randomWavelets}, we show three sets of wavelets that are computed from a dataset of Gaussian white noise. Figure~\ref{fig:randomWavelets}(a) has no variance penalty, Figure~\ref{fig:randomWavelets}(b) has a small variance penalty, Figure~\ref{fig:randomWavelets}(c) has a large variance penalty, and all wavelets are coloured according to the colour coding used in Figure~\ref{fig:waveSubspaces}(a). Despite the fact that we have used the structure of Figure~\ref{fig:waveSubspaces}(a), there is no apparent hierarchy of scales among the left set of wavelets. This highlights what we noted earlier, that the concept of scale is not built into the DDWD; rather, it must be learned from the data. When we add a small variance penalty, wavelets corresponding to finer detail subspaces are more localized, but all wavelets are jagged; this will contrast with our later examples where wavelets corresponding to later stages are smoother, reflecting the inherent structure of the later examples. Note that although the central set of wavelets was computed with non-zero variance penalty, they are not delta functions as we had asserted earlier; this is due to the dataset containing finite samples, and this effect weakens as the number of samples increases or as the variance penalty is increased (as for the right set of wavelets). In Figure~\ref{fig:randomWavelets}(c), all of the vectors are discrete delta functions: while this might seem redundant, only certain translates of the discrete delta function are included in each stage; the resulting basis consists of delta functions localized at each mesh point.

\subsection*{Kuramoto-Sivashinsky chaos}
\begin{figure}
  \begin{center}
  \includegraphics[width=0.9\linewidth]{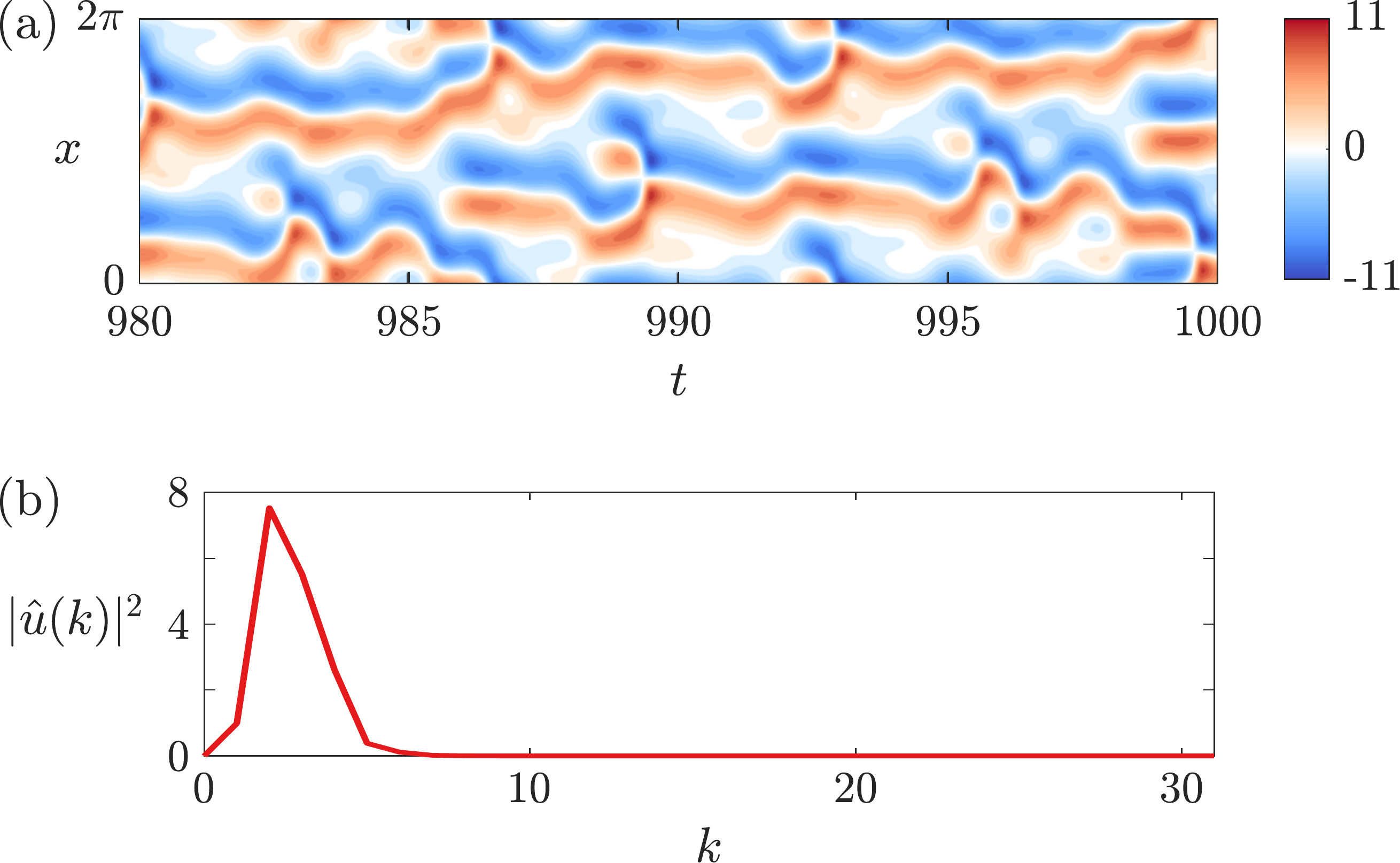}
  \end{center}
  \caption{Trajectory (a) and attendant power spectrum (b) of the Kuramoto-Sivashinsky equation.}
  \label{fig:kse}
\end{figure}

The second dataset we consider comes from the Kuramoto-Sivashinsky equation,
\begin{equation}
  \label{eq:4}
  u_t + uu_x + u_{xx} + \nu u_{xxxx} = 0
\end{equation}
for $0 \le x \le 2\pi$, with periodic boundary conditions and $\nu = (\pi / 11)^2$, which yields chaotic dynamics. We compute a numerical solution using a pseudo-spectral method with 64 Fourier modes, and assemble a dataset consisting of $90\,001$ snapshots taken from a single trajectory. The latter part of the trajectory and the power spectrum in Figure~\ref{fig:kse} clearly show that the structure is dominated by a single length scale with wavenumber $k$ around 2--3. 

We compute the DDWD with a range of variance penalties, showing the result for $\lambda^2 = 0.1$ in Figure~\ref{fig:kseWavelet} (others are shown in the Supplementary Information). We only show one set of wavelets because, no matter the variance penalty, the coarsest subspaces are the same: $V_{-6}$ is spanned by a sine wave with wavenumber $k = 2$ (the most energetic wavenumber), $W_{-6}$ is spanned by a sine wave with wavenumber $k = 3$ (the second most energetic wavenumber), and $W_{-5}$ is spanned by a vector (and its translate) containing only wavenumbers $k = 3$ and 4 ($k=4$ is the next most energetic wavenumber). The DDWD is thus robust in pushing the dominant (most energetic) length scales of the system to the lowest stages. Moreover, the energy contained in each subspace is also robust to the variance penalty (see the Supplementary Information). The first difference between wavelets computed with different variance penalties appears in the subspace $W_{-4}$, spanned by the four translates of $\psi_{-4}$. As the variance penalty is increased, the wavenumber $k = 8$ is exchanged for $k = 0$. Energetically, this makes little difference since $k = 8$ is highly damped by the hyperviscous term and contains very little energy, and $k = 0$ contains identically zero energy (for the boundary conditions we use, the spatial mean is constant and can be set to zero). The compositions of the finer detail subspaces do not change qualitatively with variance penalty, with finer detail subspaces containing higher wavenumbers. As the variance penalty is increased, localization in the Fourier domain is exchanged for localization in the spatial domain.

\begin{figure}
  \begin{center}
  \includegraphics[width=\linewidth]{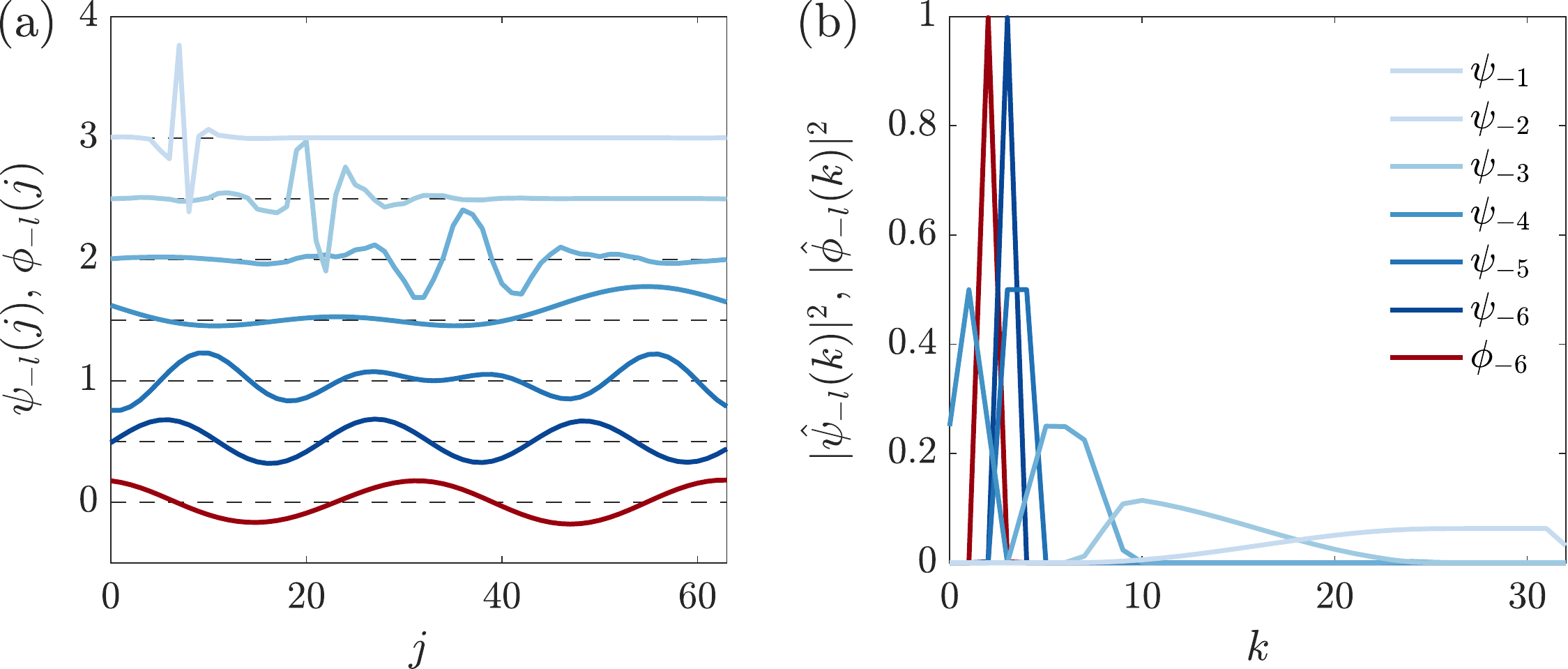}
  \end{center}
  \caption{Kuramoto-Sivashinsky wavelets (a), offset from each other by 0.5, and their power spectra (b). Colouring as in Figure~\ref{fig:waveSubspaces}(a). The variance penalty is $\lambda^2 = 0.1$. }
  \label{fig:kseWavelet}
\end{figure}

\subsection*{Homogeneous isotropic turbulence}

The final and primary dataset we consider is of forced homogeneous isotropic turbulence, taken from the Johns Hopkins Turbulence Database \footnote{{http://turbulence.pha.jhu.edu}} \citep{perlman2007data,li2008public,yeung2012dissipation}. We use a single snapshot from a direct numerical simulation on a $4096^3$ periodic grid with a Taylor-scale Reynolds number of 610.57, shown in Figure~\ref{fig:multiscale}; more details are available in the database's documentation. Our dataset consists of the velocity component aligned with 16384 randomly sampled lines (the ``longitudinal velocity'') that are parallel to the axes. Each sample is a vector of length $N = 4096$. The power spectrum is broad and has the expected $-5/3$ power law in the inertial subrange, which roughly contains wavenumbers $k \in [2, 60]$. 

Figure~\ref{fig:hitWavelet} shows the DDWD with various variance penalties (their power spectra are shown in the Supplementary Information). While at $\lambda^2= 10^{-1}$, the wavelets are well-localized only for  $l\leq 5$, for  $\lambda^2=10^0$ and $10^1$, localization is observed for $l\leq 8$ and $9$, respectively. Moreover, despite the order of magnitude difference in $\lambda^2$ between the latter two cases, the wavelets for $4\leq l\leq 8$ are nearly indistinguishable (see the Supplementary Information for more details). Furthermore, with increasing $l$, the wavelets have increasing scale: the DDWD reveals a hierarchy of scales present in the dataset, a known feature of turbulence. Recall that this feature is not built into the DDWD; rather, the method has extracted the concept of scale hierarchy from the turbulence dataset. In this case, it is appropriate to conflate stage and scale. 

It is also worth noting that with increasing variance penalty, the composition of each scale in the Fourier domain (shown in the Supplementary Information) becomes smoother and more robust, varying less across different trials. Overall, the composition of the wavelets in the Fourier domain is robust to the variance penalty. 

\begin{figure}
  \begin{center}
  \includegraphics[width=\linewidth]{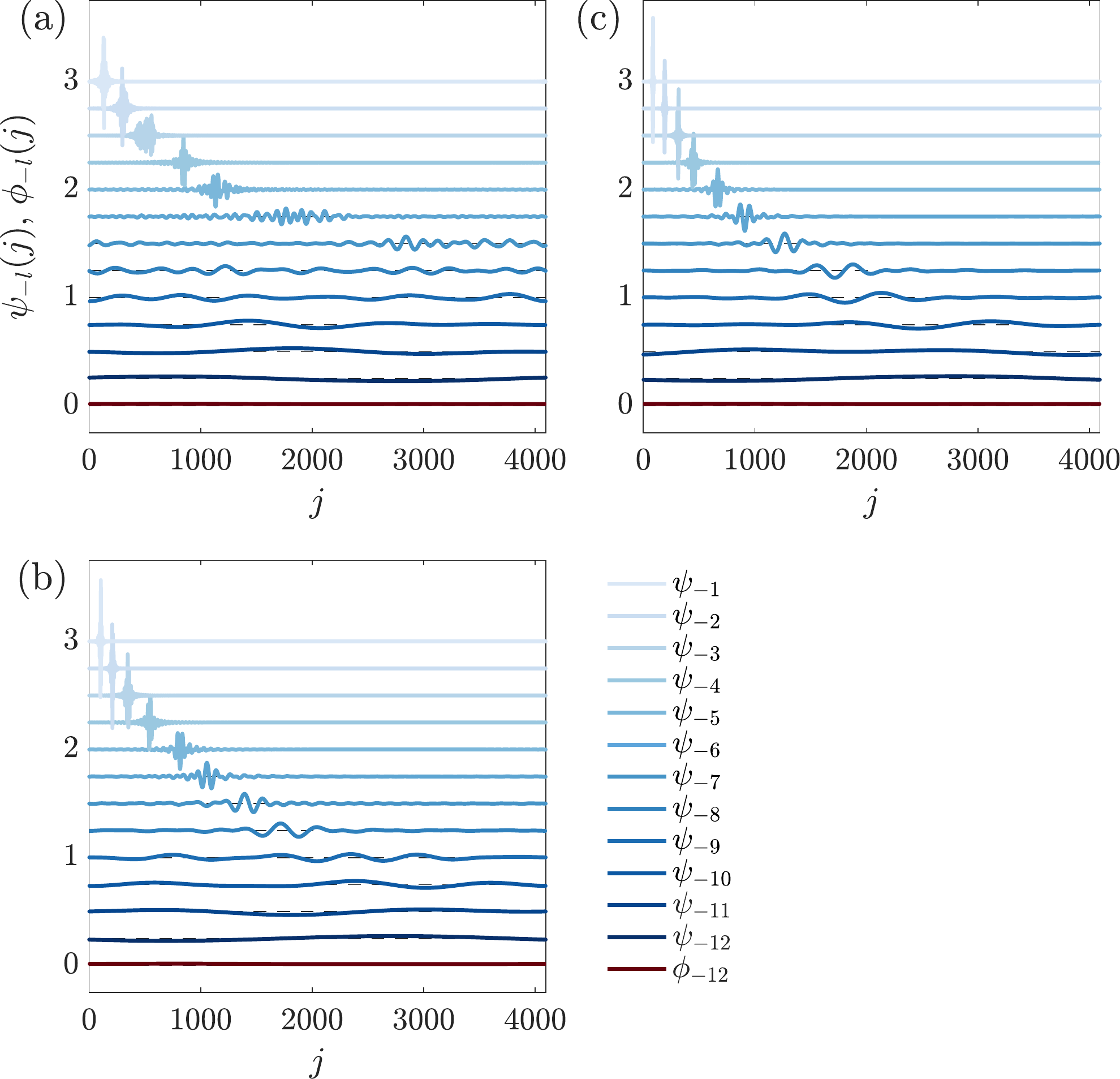}
  \end{center}
  \caption{HIT wavelets, vertically offset from each other by 0.25.  Colouring as in Figure~\ref{fig:waveSubspaces}(a). The variance penalties are $\lambda^2 = 10^{-1}$ (a), $\lambda^2 = 10^{0}$ (b), and $\lambda^2 = 10^{1}$ (c).}
  \label{fig:hitWavelet}
\end{figure}

\begin{figure}
  \begin{center}
  \includegraphics[width=\linewidth]{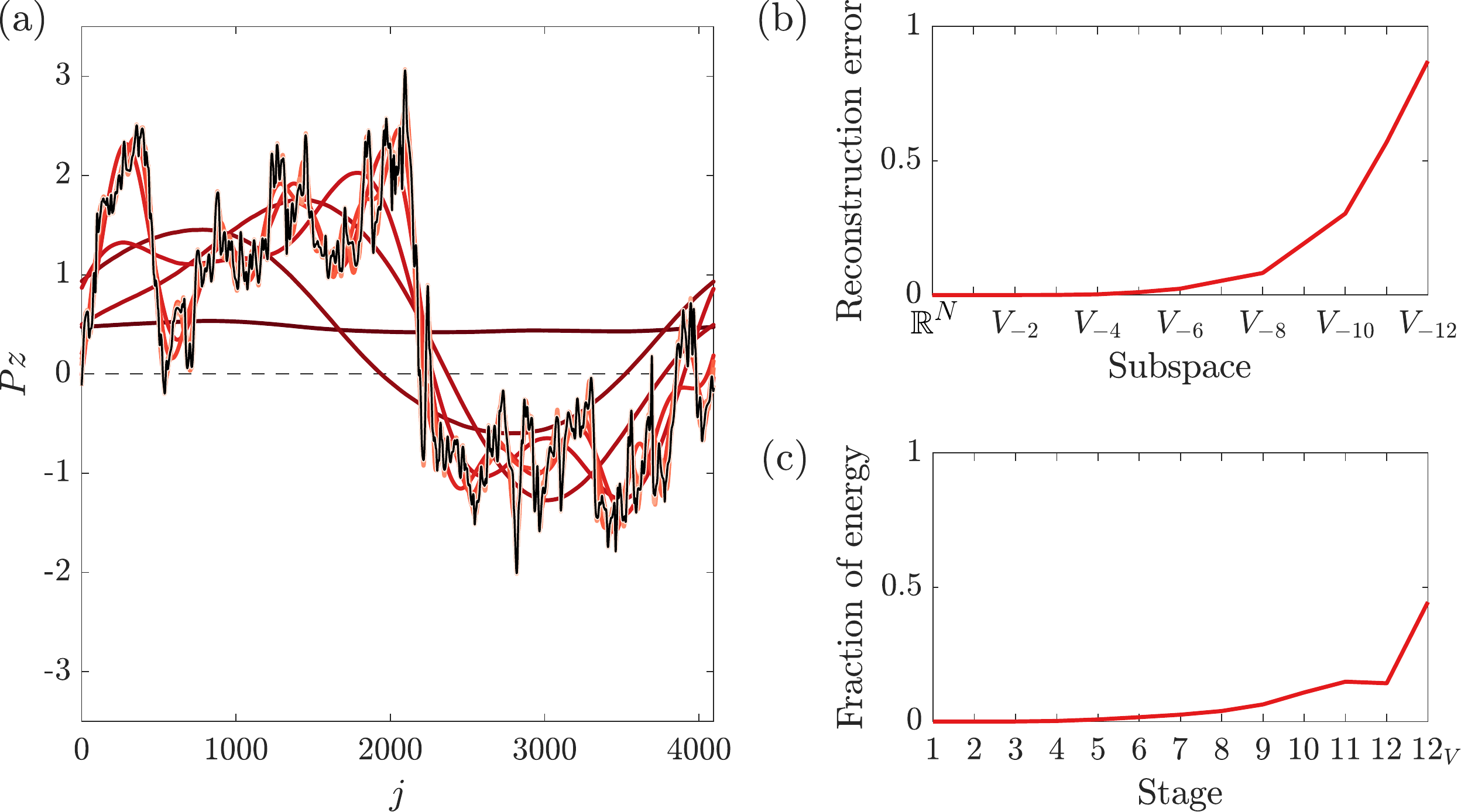}
  \end{center}
  \caption{Projection (denoted $P$) of one vector (denoted $z$) in the turbulence dataset onto the subspaces $V_{-l}$ computed with $\lambda^2 = 10^1$ (a), with colouring as in Figure~\ref{fig:waveSubspaces}(a). The thin dashed line shows the origin, and the thin solid line shows the original vector. Also shown are the reconstruction error of each projection (a), and the energy of the dataset contained in each stage for all variance penalties considered (b) ($\lambda^2 = 0, 10^{-1}, 10^0$, and $10^1$; only the result for $\lambda^2=10^1$ (red) can be seen). }
  \label{fig:hitReconstruction}
\end{figure}

To illustrate the reconstruction of data vectors using the DDWD basis,  Figure~\ref{fig:hitReconstruction}(a) shows one vector from the turbulence dataset and its projections onto the subspaces $V_{-l}$ computed with $\lambda^2=10^1$. 
Lighter colours show more detailed reconstructions and the thin black line shows the original data vector. At the coarsest level of approximation, we essentially reconstruct the spatial mean, and then add progressively finer scale features as we add smaller scale wavelet components. Figures~\ref{fig:hitReconstruction}(b) and (c), respectively,  show the reconstruction errors of the progressively finer projections, and the energy of the entire dataset contained in each stage,  for $\lambda^2=0, 10^{-1}, 10^{0}$, and $10^{1}$.   The differences in these quantities as $\lambda$ changes are visibly indistinguishable, indicating robustness of the DDWD with respect to variance penalty.

\begin{figure}
  \begin{center}	
  \includegraphics[width=\linewidth]{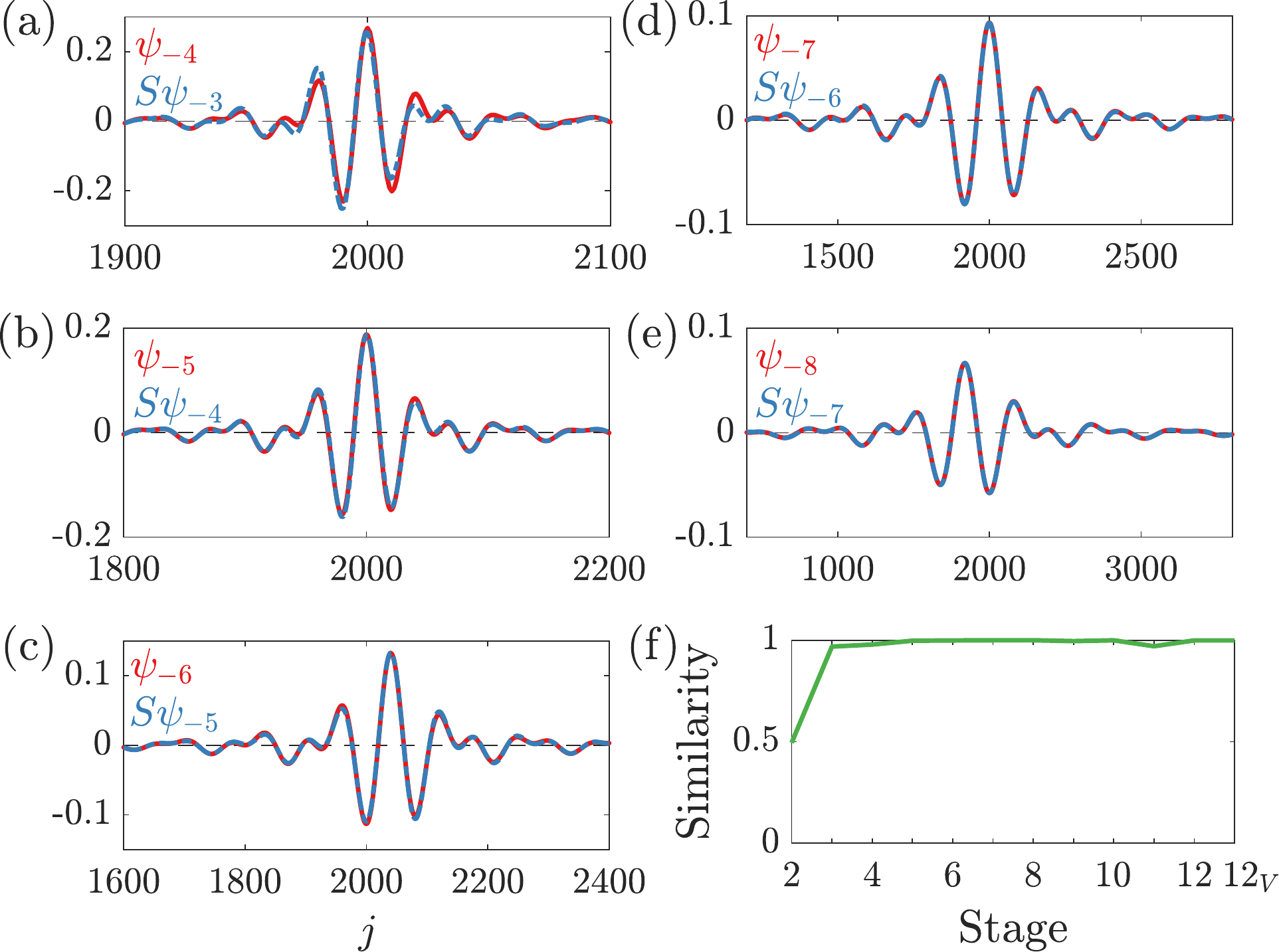}
  \end{center}
  \caption{Comparison between computed wavelets ($\lambda^2 = 10^1$) and ones obtained by dilating and rescaling the wavelet from the previous stage for stages $l = 4$ to $l = 8$ (a--e), and the level of similarity across all stages (f). }
  \label{fig:hitWaveletSim}
\end{figure}

Most interestingly, we check the wavelets that arise from the HIT data for self-similarity across stages. 
We present here results for the most localized wavelets, corresponding to $\lambda^2 = 10^1$, and show in the Supplementary Information that the same conclusions hold for $\lambda^2 = 10^0$. 
Figures~\ref{fig:hitWaveletSim}(a)--(e) show wavelets $\psi_{-l}$ for $4\leq l \leq 8$; note the change in horizontal scale from plot to plot.  Aside from their horizontal scale, these wavelets are evidently very similar looking.  The figure also shows on each plot the rescaled version of the wavelet at the previous level, $S\psi_{-l+1}$, where $S$ essentially dilates a vector by a factor of two and rescales it so that it has unit norm (see the Supplementary Information for a precise description of $S$, and for plots of  $\psi_{-l}$ and $S\psi_{-l+1}$ for all $l$.) 
For ease of comparison, we have shifted the wavelets and in some cases reflected them about their axes. In all cases shown, $\psi_{-l}$ and $S\psi_{-l+1}$ are nearly indistinguishable, indicating strong self-similarity across stages $l = 4$ to $l = 8$. This observation can be quantified:  Figure~\ref{fig:hitWaveletSim}(f) shows the inner product $\psi_{-l}^T S\psi_{-l+1}$, whose absolute value is bounded by 0 and 1, for all stages. It is very close to unity for $l>3$.
This strong self-similarity also holds for the lower variance penalty $\lambda^2 = 10^0$, as shown in the Supplementary Information, indicating that it is a robust feature derived from the data. Stages 4--8 contain the approximate wavenumbers $k \in [10,200]$, which coincides with the inertial subrange where self-similarity is expected. (The larger scales are no longer localized, so we draw no significance from the high measure of similarity in those cases.) Interestingly, the wavelets in the self-similar range are quite similar to the discrete Meyer wavelet \citep{mallat1999wavelet}, shown in Figure~\ref{fig:waveSubspaces}(b), \DFrevise{as well as to the Battle-Lemari\'e wavelet used by Meneveau in his analysis of turbulent flows \citep{meneveau1991analysis}. Performing Meneveau's analysis with our data-driven wavelets would likely yield similar results, at least in the self-similar range. }

It bears repeating that the self-similarity of the wavelets produced by the DDWD is not a result of the method, rather it is a reflection of the system. In the case of the Kuramoto-Sivashinsky system, where we know there is no similarity across scales, there is generally no relation between the data-driven wavelets across scales. For HIT, where self-similarity is hypothesized in a certain range of scales, the data-driven wavelets show self-similarity. Hellstr\"om et al.~\cite{Hellstrom:2016dw} made a somewhat related observation in turbulent pipe flow. They performed PCA on a set of experimentally obtained velocity fields from a cross-section of the pipe, and found that they could rescale the modes so that they overlapped. This observation is consistent with the attached eddy hypothesis about the structure of wall-turbulence \cite{Marusic:2019fy,Hwang:2015iz}.
 Their modes were global in space, as usually results from PCA; this is particularly true for the azimuthal direction, for which PCA yields Fourier modes due to periodicity. For the HIT data, which is periodic in all three directions, PCA would yield Fourier modes in all three directions, revealing no information about the system that could not be obtained from Fourier decomposition.

\section*{Conclusions}

We have introduced a method that integrates key aspects of PCA and wavelet analysis to yield a data-driven wavelet decomposition.
This method takes an ensemble of data vectors corresponding to field values at a lattice of points in space (or time) and generates a hierarchical orthogonal basis.
In contrast to traditional wavelet bases, the basis elements at each stage are not simply dilations of given mother or father wavelets, but rather are determined stage-by-stage from the data.  
For data that is not self-similar, neither are the resulting  basis elements. Rather, these represent the differing structures at the different stages.
In contrast, for self-similar data, the basis vectors at different stages are related to one another by a simple rescaling.
Indeed, for data from homogeneous isotropic turbulence---a high-dimensional, nonlinear, multiscale process---we show self-similarity of the wavelet basis elements, which in turn reveals the self-similarity of the data, providing direct evidence for a century-old phenomenological picture of turbulence. 
 
Future work on the DDWD will need to extend the methodology to multiple dimensions, different boundary conditions, and unstructured domains. As a start, tensor products can be used to address the first issue, boundary wavelets can be used to address the second issue \cite{mallat1999wavelet}, and wavelets on graphs can be used to address the last issue \cite{hammond2011wavelets}. \DFrevise{For incompressible fluid flows, velocity fields are vector-valued and divergence-free; Farge et al.~\cite{farge2003coherent} provides a few options to handle this case that may be generalizable to the data-driven case.} Attention must also be given to the development of efficient optimization algorithms for computing the basis.
Finally, based on the ability of the present method to extract self-similar basis elements from self-similar turbulent flow data, we view it as a potentially important new starting point for identification and characterization of localized hierarchical turbulent structures in a wide variety of fluid flows, as well as other complex multiscale systems. \DFrevise{We are particularly interested in applying the DDWD to wall-bounded flows and making connections with the attached eddy model of turbulence. }

\acknow{This work was supported by AFOSR grant FA9550-18-0174 and ONR grant N00014-18-1-2865 (Vannevar Bush Faculty Fellowship). }

\showacknow{} 

\bibliography{references,turbulenceMDG,miscMDG}

\end{document}